\documentclass[lettersize,journal]{IEEEtran}
\usepackage{amsmath,amsfonts}
\usepackage{algorithmic}
\usepackage{algorithm}
\usepackage{array}
\usepackage[caption=false,font=normalsize,labelfont=sf,textfont=sf]{subfig}
\usepackage{textcomp}
\usepackage{stfloats}
\usepackage{url}
\usepackage{verbatim}
\usepackage{graphicx}
\usepackage{multirow}
\usepackage{tablefootnote}
\usepackage{soul}
\usepackage{cite}
\usepackage{threeparttable}

\begin{document}

\title{Generative Motion In-betweening by Diffusion over Continuous Implicit Representations}


\author{Shiyu Fan, Paul Henderson, Edmond S. L. Ho
\thanks{All authors are with the School of Computing Science, University of Glasgow, United Kingdom. E-mails: s.fan.1@research.gla.ac.uk, Paul.Henderson@glasgow.ac.uk, and Shu-Lim.Ho@glasgow.ac.uk}
\thanks{Edmond S. L. Ho is the corresponding author}
}

\maketitle

\begin{figure*}[!t]
\includegraphics[width=\linewidth]{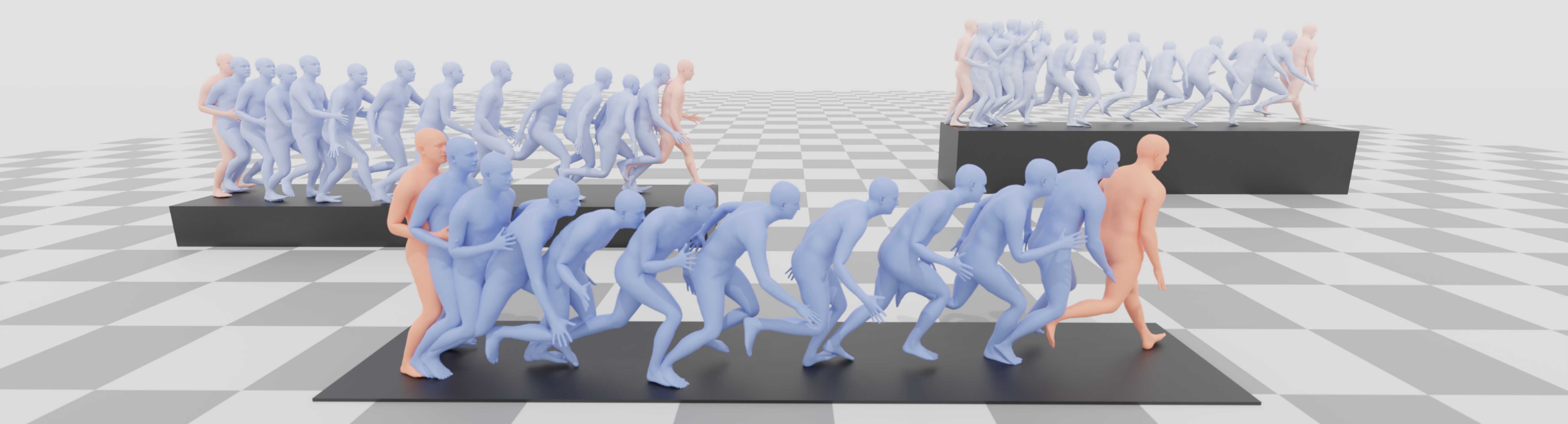}
  \caption{Given the same set of initial and final keyframes, our model generates a diverse range of in-between motions.}
  \label{fig:teaser}
\end{figure*}

\begin{abstract}
Recent advances in generative models have yielded impressive progress on motion in-betweening, allowing for more complex, varied, and realistic motion transitions. However, recent methods still exhibit noticeable limitations in preserving keyframe information and ensuring motion continuity. 
In this paper, we propose a novel pipeline and sampling optimization strategy for latent diffusion models (LDM) based on motion implicit neural representations (INR). By establishing a mapping between INR and sparse spatial or temporal information within latent diffusion, our model can sample the INR parameters from extremely sparse and ambiguous keyframe data and reconstruct plausible and smooth motions from the manifold. Our experiments demonstrate the superior performance of our model, which significantly improves motion generation quality in scenarios with few keyframes while ensuring both keyframe accuracy and diversity of in-between motions.
\end{abstract}

\maketitle
\section{Introduction}
\IEEEPARstart{A}{s} a fundamental task of character animation, motion in-betweening involves generating smooth and natural transitions between keyframes in a motion sequence. One of the key challenges is to satisfy the spatial and temporal constraints specified in the keyframes (i.e. target poses) while minimizing unnatural movements like jittering and foot sliding. With the aid of motion in-betweening algorithms, high-quality motions can be generated with only several sparse keyframes as in the examples shown in  Fig.~\ref{fig:teaser}.

Most recently, motion in-betweening has evolved from a regression task \cite{Harvey:SIGGRAPH2020} to a generative task. The emergence of generative models, notably diffusion models \cite{ho2020denoising}, has brought new life to the traditional motion in-betweening task. Diffusion-based methods have demonstrated promising capabilities for generating diverse and realistic human motions \cite{tevet2023human,Zhang/TPAMI.2024.3355414}. Some diffusion models incorporate motion editing from text-to-motion tasks for motion in-betweening~\cite{xie2024omnicontrol,DNO_2024_CVPR}.

However, existing diffusion models also present significant challenges, which include slow training and generation speeds and wide disparities between sampling results and keyframes. 
Another limitation of the prior motion in-betweening algorithms~\cite{Cohan:SIGGRAPH2024:CondMDI,NeMF} is low generation quality 
when the keyframes are extremely sparse. To tackle this problem, models such as CondMDI~\cite{Cohan:SIGGRAPH2024:CondMDI} require a text description in addition to the keyframes in order to generate high-quality in-between motions. 
Without such additional input, the results may contain unnatural movement, such as foot sliding or stopping midway through a motion. Such a dependency limits the practicality of the method in scenarios where text conditioning is not available. Additionally, it exhibits large keyframe errors.

In order to generate realistic in-between motion 
from sparse spatial-temporal information, it is essential to utilize a continuous representation to capture global information. 
By constructing compact and continuous functions, implicit neural representation (INR) captures the global representation of motion sequences and makes it possible to generate smooth and natural motion transitions. 
Since the encoded motion representation no longer contains explicit spatial or temporal features, inferring the corresponding INR parameters from sparse keyframes is highly challenging.
This inspires us to propose a novel latent diffusion sampler for motion INR.  
Specifically, our model establishes a correlation between randomly selected keyframes and INR latent by the cross-attention module in the denoising network, allowing for INR manifold sampling conditioned on keyframes. To further improve the motion quality and keyframe error, we introduce the concept of manifold preservation for the sampling optimizations and incorporate additional constraints tailored to motion INR. This optimization module, namely Implicit Manifold Guidance (IMG), corrects the sampling direction during the sampling process by minimizing the geometric error and manifold error. 

Relying on the capability of the implicit neural representation, our model can successfully sample smooth and realistic motion given only sparse, randomly located keyframes. 
Furthermore, our approach produces motions with reduced keyframe error and which exhibits greater variability. 
The contributions of this work include:
\begin{itemize}
    \item We are the first to adapt Motion INR to the generative motion in-betweening task.
    To achieve this, we propose an INR manifold sampler based on latent diffusion model (LDM) within the implicit neural field, which leverages compact continuous-time motion representations to generate realistic and high-quality motion conditioned on extremely sparse keyframes.
    \item We introduce the idea of manifold preservation and design a novel loss in the posterior sampling optimization algorithm tailored for the motion implicit manifold to preserve motion quality while reducing keyframe error. 
\end{itemize}
Our method achieves state-of-the-art performance in motion in-betweening tasks conditioned on extremely sparse keyframes without textual descriptions. \textit{Explore codes at: https://github.com/Coondinator/NINB}

\section{Related Works}

\subsection{Motion In-betweening}
Prior to the emergence of deep learning, the majority of work~\cite{rose2001artist, Ciccone} relied on kinematics interpolation to address this problem.
Rigid body physics simulation~\cite{Tessler:ToG2024, Gopinath:SIGGRAPH2022poster}
and motion imitation~\cite{Li:ACM-MM2022} were used
for generating plausible in-between frames.

With the rise of neural networks, time series models have been applied to the motion prediction task~\cite{10.1145/3283254.3283277,Gopalakrishnan:CVPR2019,Tao:ECCV2020,Kim:PR2022, Harvey:SIGGRAPH2020,Oreshkin:TVCG2023}. Qin et al.~\cite{Qin:SA2022} stack multi-stage Transformers to progressively generate in-between motion. AvatarGPT~\cite{zhou2024avatargpt} encodes motions as tokens and uses a powerful large language model to achieve both generation and in-betweening tasks. Meanwhile, the concept of high-dimensional manifolds has been gradually adopted in motion encoding and generation~\cite{holden2015learning}. DeepPhase~\cite{DeepPhase} presents the periodic phase of the motion manifold, enabling the encoder to capture the periodicity of motions in time. This manifold can also serve as a prior for motion in-betweening~\cite{Starke:2023, Hong:CGF2024}. Although those auto-regressive methods can provide details by learning from a large dataset, they lack diversity in the generated results. 

Therefore, there is a growing desire that generative models can bring more diversity to motion synthesis and in-betweening. Petrovich et al.~\cite{petrovich2021action} apply a Variational autoencoder (VAE) to motion synthesis. Ren et al.~\cite{Ren:TVCG2025} synthesize forward and backward motion sequences by introducing a bi-directional stitching CVAE. Due to their superior generative performance, diffusion models have gradually replaced GANs and VAEs. CondMDI~\cite{Cohan:SIGGRAPH2024:CondMDI} incorporated the diffusion model to handle different types of in-betweening tasks. However, its weak condition constraint results in significant keyframe errors, and learning from discrete-time sequences often leads to abrupt motion transitions. Similar to text-to-motion models, it also relies on text condition inputs to improve motion quality.

\subsection{Motion Diffusion Model and Motion Editing}
Given the superior performance of diffusion models in image generation tasks, this leading generative model has gradually been applied to the field of motion generation. MDM~\cite{tevet2023human} and MotionDiffuse~\cite{Zhang/TPAMI.2024.3355414} were the first two trials introducing diffusion models into motion generation, and significantly improved the performance of text-to-motion synthesis. Motion latent diffusion~\cite{mld} encodes motions into a latent space and performs the diffusion process in this space. This design greatly shortens the training and inference time.
Diffusion models have also been introduced into the generation of two-person~\cite{PriorMDM,liang2024intergen} and further extended to multi-person interactions~\cite{Xu:Multi-person}. 

Another challenge is how to control the generated results from diffusion. Results generated solely based on prompts may not meet other requirements, such as temporal and geometric constraints (e.g. when and where the joints should be located in certain keyframes). This issue can be addressed in two ways: 1. actively intervene in the denoiser by guiding and editing the generation process through adding embeddings with constraint information; or 2. manipulating the attention mechanism as in ControlNet~\cite{ControlNet} and Prompt-to-Prompt~\cite{hertz2022prompt}. For motion tasks, OmniControl~\cite{xie2024omnicontrol} constrains motion embeddings through an additional auxiliary network for motion editing. 
Such solutions, though effective, often require redesigning network components and incur additional training costs. 
MoMo~\cite{MOMO} achieved zero-shot motion style transfer by combining the Query, Key, and Value from the self-attention module of different characters during the denoising process. 
DNO~\cite{DNO_2024_CVPR} also attempts to guide the sampling without changing the denoiser structure. 
For a given criterion (like keyframe poses), DNO computes the error based on the sampling results after each ODE solver's sampling process, then updates the initial noise or inverted noise through gradient backpropagation.
However, this algorithm requires full sampling at each optimization step, greatly increasing computational costs and memory usage. To remain feasible, it drastically reduces the number of diffusion steps, leading to a complete loss of diversity. 
\section{Why Implicit Neural Representations?}
\label{sec:pre}
Without providing context through text conditioning, time-discrete modeling tends to lose the horizon between critical information across long-range dependencies and introduces noise. This leads to discontinuities and jittering motions, especially in in-betweening with sparse keyframes. 

INRs represent data implicitly as a function that maps continuous parameters to corresponding data values. Such representations have been widely applied to 3D scene representation and image upsampling \cite{mildenhall2021nerf, gao2023implicit,chen2021learning}. Motion INR, in particular, represents motion by a neural network $\mathbb{D}(\cdot)$ that maps from a continuous time dimension to continuously-varying poses, typically conditioned on a latent vector $z$ which determines the specific motion captured.  According to the given time sequence $\left \{   t_{k}\right \}_{k=0}^T$, the INR decoder $\mathbb{D} \,:\, (z, t)\mapsto (X_{t},R_{t})$ independently decodes the latent $z$ to the root transform $X_{t}$ and joint rotation $R_{t}$ at different times; these are then assembled into a complete and smooth motion.

Given a discretely-sampled motion sequence $m\in \mathbb{R}^{T\times D}$, an encoder $\mathbb{E}(\cdot)$ may be used to obtain the corresponding latent $z=\mathbb{E}(m)$; this is used to train the model via an autoencoding loss. The comparisons between INR and other representations are given in (Sec.~\ref {sec:abla_inr}). 

In motion in-betweening, the use of INR offers the following advantages: 
\begin{itemize}
\item{Encoded from the whole motion sequence by a VAE, INRs can effectively capture and preserve the motion patterns even without text conditions, consistent with the Matching Score and R-precision results reported in Table~\ref{tab:random_key} and Table~\ref{tab:content_key}.}

\item{By modeling motion as a continuous function over time, the generated motions from INR exhibit smoother and more natural transitions compared to other motion representations (as shown in Sec~\ref{sec:ran_key_exp}).}

\item{The results in Table~\ref{tab:cross_dataset} show that sampling within the INR space generalizes well to unseen conditions, suggesting that the INR representation provides a smooth and transferable manifold for motion generation.}
\item{Due to the compression of the time dimension, INRs significantly reduce the computational complexity of diffusion sampling, making it convenient for efficient sampling guidance (Fig.~\ref{fig:cal}).}

\end{itemize}
Some previous work \cite{NeMF} has tried to utilize INR for motion synthesis. However, they also suffer from significant limitations. It requires reparameterization using more detailed initial motions, and gradient descent is performed on the reparameterized latent to fit the keyframes. This causes its sampling to degrade into a deterministic process, and under the condition of sparse keyframes, it falls into a local minimum and significantly loses motion details. By training a diffusion-based sampler over the INR's latent space $z$, our model avoids the aforementioned issues. It ensures high-quality and diverse motion generation while significantly reducing keyframe errors.

\begin{figure*}[htb]
  \centering
  \includegraphics[width=\textwidth]{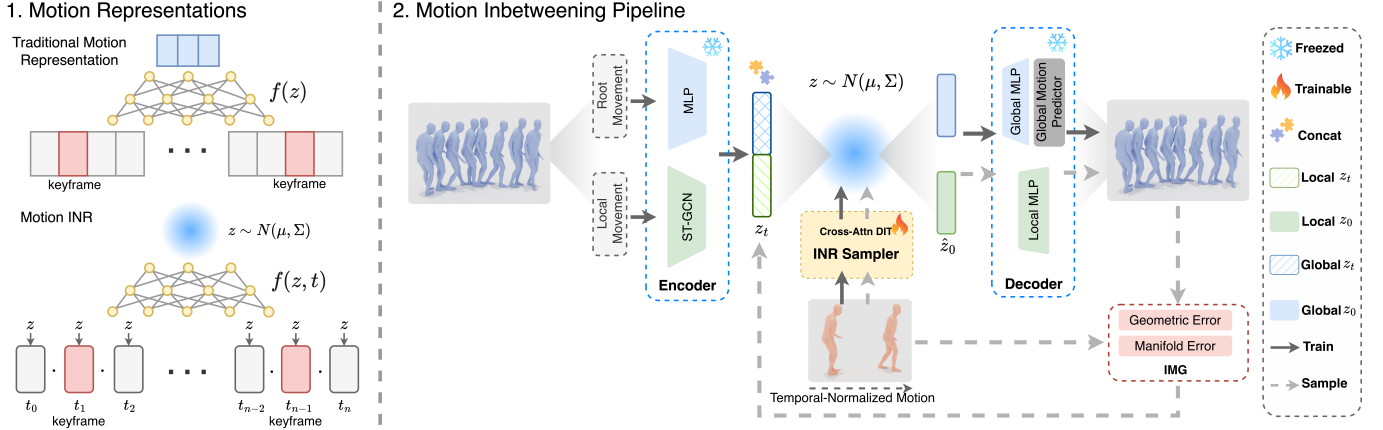}
  \caption{(1). The differences between common motion representations and INR over the motion inbetweening task. (2). The overview of our proposed motion in-betweening pipeline. \textbf{Training Stage:} The LDM denoises the noisy latent by the random keyframes extracted from the motion sequence.  \textbf{Sampling Stage:} During sampling, the module of implicit manifold guidance calculates the geometric error and manifold errors to optimize the posterior estimation.}
  \centering
  \label{fig:pipeline}
\end{figure*}
 
\section{Methodology} \label{sec:met}
Our objective is to sample plausible motions given sparse keyframe conditioning. We cast this as finding parameters in the INR manifold that align with the spatial constraints imposed by the keyframes. Specifically, we propose a latent diffusion pipeline over the latent space of NeMF~\cite{NeMF}. The details of this diffusion model are shown in Sec.~\ref{sec:ldm}. 
We further propose a custom inference guidance for preserving keyframe content (Sec.~\ref{sec:img}).

\subsection{Motion Representation} \label{sec:mot_rep}
We represent each motion sequence $m$ as two parts: global and local. The global part contains the time-varying root orientation $\mathbf{R}_{o} \in \mathbb{R}^{T\times 6}$ and root translation $\mathbf{R}_{p} \in \mathbb{R}^{T\times 3}$ in world coordinates, while the local part represents motion data of each joint with the position 
$\mathbf{X}_{p} \in \mathbb{R}^{T\times J \times 3}$, velocities $\mathbf{X}_{v} \in \mathbb{R}^{T\times J \times 3}$, rotations $\mathbf{X}_{r} \in \mathbb{R}^{T\times J \times 6}$ in 6D rotation form~\cite{zhou2019continuity}, and angular velocities $\mathbf{X}_{a} \in \mathbb{R}^{T\times J \times 3}$.
Here $T$ denotes the length of the motion sequence and $J$ is the number of joints. 

\subsection{Sampler for INR} \label{sec:ldm}

\subsubsection{Model Structure}
Our pipeline consists of two main components: a motion INR VAE and a latent diffusion sampler. The VAE is built upon NeMF’s structure and pretrained parameters. During both training and sampling (Fig.~\ref{fig:pipeline}), the VAE remains frozen. 
To model the relationship between keyframes and the manifold learnt by the VAE, we build a sampler with a backbone based on diffusion transformer (DiT)~\cite{DiT:ICCV2023}. We replace DiT's self-attention with cross-attention over the keyframes to better fit our task. Transformer-based denoising networks effectively integrate information from sparse keyframes, helping map them to our compact latent representations despite large temporal gaps.
Our diffusion network is composed of four transformer blocks, incorporating adaLN-Zero layers that transform activations based on the diffusion timestep.

The motion is decomposed into local and global data (Sec.~\ref{sec:mot_rep}), and the encoder correspondingly maps to a local latent $z_{l} \in \mathbb{R}^{1024}$ and a global latent $z_{g} \in \mathbb{R}^{256}$. To reconstruct the motion, $z_{l}$ and $z_{g}$ are concatenated into a single 1280-D vector $z_{0}$ as the input of the denoiser.

To enable conditioning on arbitrary keyframes using a fixed-length representation, we first perform positional encoding on the original motion frames, then mask these to retain only the keyframes, replacing the remainder with zeros. 
The keyframe embeddings are then used to derive keys and values in the cross-attention blocks. 

\subsubsection{Training}
Following DDPM\cite{ho2020denoising}, we train a denoiser $\mathit{DM}(z_{t}, t)$ to predict $z_{0}$ and calculate the Mean Squared Error of it with the ground truth, where $z_{t}$ is the noisy latent at the diffusion forward step $t$:
\begin{equation}
\mathcal{L}_{z} := \mathbb{E}_{z_{0}, z_{t}} \left\|z_{0}-\mathit{DM}(z_{t}, t)\right\|^{2}
\end{equation}
To achieve flexible motion in-betweening from random keyframes, we choose the conditioning keyframes randomly from the ground-truth motion, by independent Bernoulli sampling.

Given the importance of minimizing keyframe error
in the in-betweening task, we introduce the geometric consistency terms to the objective function. 
The predicted $\hat{z}_{0}=\mathit{DM}(z_{t}, t)$ is decoded into motion $\hat{x}$ through the decoder and its errors are calculated at the keyframes, including root translation error $\mathcal{L}_{trans}$ and local joint translation error $\mathcal{L}_{pos}$ which are computed based on the mean absolute error. The $x_{trans}$ and $x_{pos}$ represent the root translation and local joint position of ground truth, and $\hat{x}_{trans}$ and $\hat{x}_{pos}$ are predicted root translation and local joint position:
\begin{equation}
\mathcal{L}_{trans} =  \frac{1}{k}\sum^{k}\left\|\hat{x}_{trans}-x_{trans}\right\|_{1}
\end{equation}
\begin{equation}
\mathcal{L}_{pos} =\frac{1}{k}\sum^{k} \left \|\hat{x}_{pos}-x_{pos}\right \|_{1}
\end{equation}
Root orientation error $\mathcal{L}_{ori}$ and local joint rotation error $\mathcal{L}_{rot}$ are calculated using geodesic distance:
\begin{equation}
\mathcal{L}_{ori} = \frac{1}{k}\sum^{k} \arccos \frac{Tr(\mathbf{R}_{t}^{o}(\hat{\mathbf{R}}_{t}^{o})^{-1})-1}{2}
\end{equation}
\begin{equation}
\mathcal{L}_{rot}=\frac{1}{k}\sum^{k} \arccos \frac{Tr(\mathbf{R}_{t}(\hat{\mathbf{R}}_{t})^{-1})-1}{2}
\end{equation}
where $k$ is the number of keyframes for each motion sequence, $\mathbf{R}_{t}$ and $\mathbf{R}_{t}^{o}$ are the ground-truth joint rotation matrix and root orientation matrix of the keyframes, $\hat{\mathbf{R}}_{t}$ and $\hat{\mathbf{R}}_{t}^{o}$ are the predicted joint rotation matrix and root orientation matrix, respectively. Hence, the final training objective function of our diffusion model is: 
\begin{equation}
\begin{split}
\mathcal{L}_{total} & = \lambda_{z}\mathcal{L}_{z} + \lambda_{pos}\mathcal{L}_{pos} 
+\lambda_{trans}\mathcal{L}_{trans} +\lambda_{rot}\mathcal{L}_{rot} \\ &+  \lambda_{ori}\mathcal{L}_{ori}
\end{split}
\end{equation}
where $\lambda_{z}$, $\lambda_{pos}$, $\lambda_{trans}$, $\lambda_{ori}$ and $\lambda_{rot}$ are the loss scales for each term. During training, we randomly drop the input conditionings with a probability of 10\% to enable Classifier-free guidance (CFG) during sampling~\cite{cfg}.

\subsection{Implicit Manifold Guidance (IMG)}\label{sec:img}

\begin{figure}[h]
  \centering
  \includegraphics[width=1\linewidth]{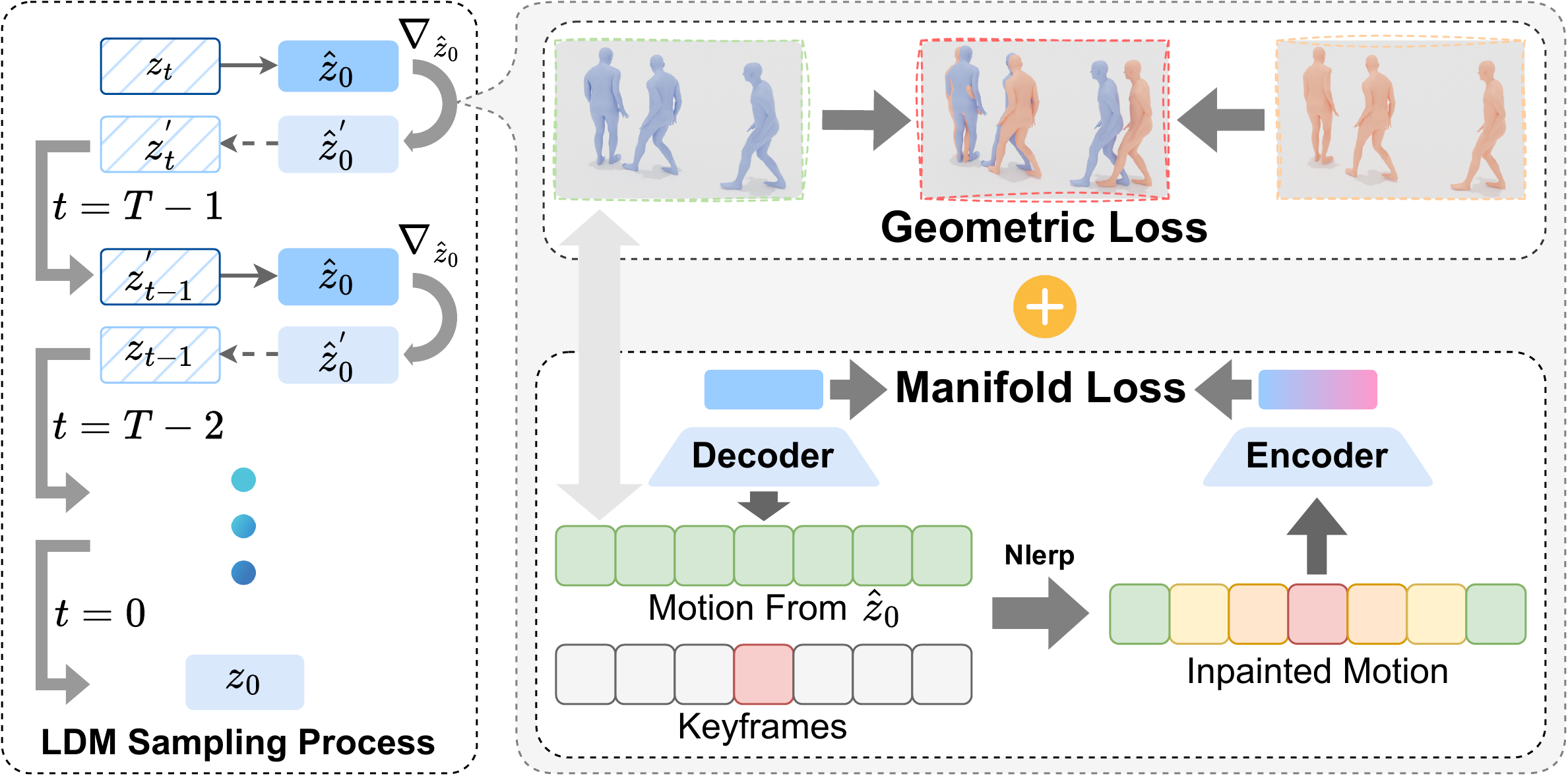}
  \caption{The implementation details of IMG: The optimization is conducted at each sampling step. The geometric and manifold errors computed from the posterior estimation $\hat{z}_{0}$ determine the optimization gradient direction for $\hat{z}_{0}$. It thereby refines the overall sampling direction.
  }
  \vspace{-0.5em}
  \label{fig:IMG}
\end{figure}

While conditioning with CFG is theoretically sufficient for the model that has been trained with geometric consistency loss, in practice, the randomness of the diffusion model and reconstruction error may still lead the generated output to diverge from the keyframes. We therefore propose a novel guidance strategy designed to improve adherence to the keyframes, while keeping motion diverse and realistic, namely \textbf{Implicit Manifold Guidance (IMG)} (Fig.~\ref{fig:IMG}). 

For the stochastic differential equation (SDE) diffusion model (e.g. DDIM \cite{songdenoising}), the sampling process from noisy data $x_{t}$ to $x_{t-1}$ can be regarded as computing the conditional probability distribution $q(x_{t}|x_{t-1}, x_{0})$ given timestep $t \in[0, T]$ and clean data $x_{0}$. With the cumulative product $\bar{\alpha}_{t}$ of noise coefficient $\alpha_{t}$ and variance $\sigma_{t}$ of the distribution $q(x_{t}|x_{t-1}, x_{0})$, sampling of $x_{t}$ to $x_{t-1}$ is done by:
\begin{equation}
\begin{split}
    x_{t-1}=   
    \sqrt{\bar{\alpha}_{t-1}}\hat{x}_{0}+\sqrt{1- \bar{\alpha}_{t-1} - \sigma_{t}^{2}}\cdot \frac{x_{t}-\sqrt{\bar{\alpha}_{t}}\hat{x}_{0}}{\sqrt{1-\bar{a}_{t}}}+ \sigma_{t}\epsilon_{t}^{*}
\end{split}
\end{equation}

To solve inverse problems, diffusion posterior sampling  (DPS) \cite{chung2023diffusion} updates the forward direction of $x_{t-1}$ at each sampling step based on the error $L$ between the measurement $\mathcal{A}(\hat{x}_{0|t})$ of the posterior sampling result $\hat{x}_{0|t}$ from the Tweedie’s
formula and the observed value $y$:
\begin{equation}
    x_{t-1} \gets x_{t-1} - \lambda_{i}\nabla_{x_{t}} L(\hat{x}_{0|t};y)
\end{equation}

Although the optimization algorithms used in DPS can bring the generated results closer to the observed values, i.e. the keyframes in our task, their optimization performance is highly related to the gradient coefficients and sampling timesteps. As a result, achieving optimal optimization requires smaller scales and more iterative steps. Moreover, for implicit manifolds, performing gradient optimization based on the denoised results at each step $z_{t}$ can cause the generated manifold to deviate from the original manifold distribution.
To tackle these problems, we introduce the idea of manifold-preserving guidance (MPGD)~\cite{he2024manifold}.
It applies gradient descent on posterior estimation $z_{0}$, which accelerates convergence compared to DPS.

\begin{algorithm}[H]

    \caption{Implicit Manifold Guidance}
    \label{alg:ours}

        $\textbf{Input:}\ T,  y, m, \lambda_{i=1}^{T}, \gamma_{i=1}^{T}, \left \{  \tilde{\sigma_{i}}   \right \}_{i=1}^{T},  \mathcal{E},  \mathbb{D}, \mathcal{A},  \ z_{T}\sim \mathcal{N}(\mathbf{0}, I)$

        \textbf{For} {$i = T$ \textbf{To} 0}
        
        \hspace{0.3cm} $\hat{s}\gets s_{\theta}(z_{i}, i), $
        $\hat{z}_{0} \gets \frac{1}{\sqrt{\bar{\alpha}_{i}}} (z_{i}+(1-\bar{\alpha}_{i})\hat{s})$
               
        \hspace{0.3cm}$ \epsilon \sim \mathcal{N}(\mathbf{0}, I)$

        \hspace{0.3cm} \textbf{If}\ {$ i \le 300$}{
        
        \hspace{0.3cm}\hspace{0.5cm}$\hat{z}_{inp}\gets \mathcal{E}( (\mathit{m}w_{k})*Slerp(y)+(1-(\mathit{m}w_{k}))* \mathbb{D}(\hat{z}_{0}))$
        
        \hspace{0.3cm}\hspace{0.5cm}$\hat{z}_{0}^{'} \gets\hat{z}_{0} - \lambda_{i}\nabla_{\hat{z}_{0}}\mathcal{L}_{geo}(\mathbf{y}, \mathcal{A}(\mathbb{D}(\hat{z}_{0})))$
        
        \hspace{0.3cm}\hspace{0.5cm}$\hat{z}_{0}^{'} \gets\hat{z}_{0}^{'} - \gamma_{i}\nabla_{\hat{z}_{0}} \left \| \hat{z}_{0}-\hat{z}_{inp} \right \|_{2}^{2}$

        \hspace{0.3cm}\hspace{0.5cm}$z_{i-1} \gets \frac{\sqrt{\alpha_{i}}(1-\bar{\alpha}_{i-1}) }{1-\bar{\alpha}_{i}} z_{i}+\frac{\sqrt{\alpha_{i-1}}\beta_{i} }{1-\bar{\alpha}_{i}}\hat{z}_{0}^{'}+ \hat{\sigma}_{i}\epsilon$
        }
        
        \hspace{0.3cm}\textbf{Else}
        
        \hspace{0.3cm}\hspace{0.5cm}{$z_{i-1} \gets \frac{\sqrt{\alpha_{i}}(1-\bar{\alpha}_{i-1}) }{1-\bar{\alpha}_{i}} z_{i}+\frac{\sqrt{\alpha_{i-1}}\beta_{i} }{1-\bar{\alpha}_{i}}\hat{z}_{0}+ \hat{\sigma}_{i}\epsilon$}

        $\textbf{Return:}\ z_{0}$
  
\end{algorithm}

Following MPGD, our IMG employs a geometric loss for guidance, which is the same as the settings in the training stage. 
However, directly performing gradient descent in Euclidean space $\nabla_{\hat{z}_0}\mathcal{L}_{geo}$ can cause the results to deviate from the manifold surface mapped from real data. Therefore, we additionally propose a manifold correction term to further guide the gradient descent direction of $z_{0}$.  For each step, the specific optimization process of IMG is as follows.
After obtain the posterior estimate $\hat{z}_{0}$ from denoiser, the INR decoder $\mathbb{D}(\cdot)$ decodes the $\hat{z}_{0}$ into motion.
The geometric errors between the generated motion and ground truth keyframe values $y$  will be calculated, including the errors on root translation, root orientation, local joint position, and joint rotation: 
\begin{equation}
\begin{split}
\mathcal{L}_{geo}(y, \mathcal{A}(\mathbb{D}(\hat{z}_{0})))& = \lambda_{pos}\mathcal{L}_{pos} +\lambda_{trans}\mathcal{L}_{trans} \\ & + \lambda_{rot}\mathcal{L}_{rot}+  \lambda_{ori}\mathcal{L}_{ori}
\end{split}
\end{equation}
For the manifold error, an initial reference motion $Slerp(y)$ is generated using the given keyframes $y$ by spherical linear interpolation. Based on the keyframe mask $m$, we construct a Gaussian kernel $w_{k}$ and use this to smooth the keyframe weights. The motion decoded from the predicted $\hat{z}_{0}$ is linearly interpolated with $Slerp(y)$ by the keyframe weights. The interpolated result is then fed back into the encoder $\mathcal{E}(\cdot)$ to obtain an inpainted latent representation $\hat{z}_{inp}$ of the motion. We compute the error of this new latent with respect to the originally predicted $\hat{z}_{0}$, and then perform gradient descent on $\hat{z}_{0}$. The update formula for $\hat{z}_{0}$ is as follows: 
\begin{equation}
\begin{split}
\hat{z}_{0}^{'}\gets\hat{z}_{0}- \lambda_{i}\nabla_{\hat{z}_{0}}\mathcal{L}_{geo}-\gamma_{i}\nabla_{\hat{z}_{0}} \left \| \hat{z}_{0}-\hat{z}_{inp} \right \|_{2}^{2}
\end{split}
\end{equation}
where $\lambda_{i}$ and $\gamma_{i}$ are the coefficients of the gradient of the geometric term and manifold term, respectively. Subsequently, $z_{t-1}$ is updated according to the optimized $\hat{z}_{0}^{'}$.
We choose the classic DDIM solver with 1000 sampling steps.
We empirically apply IMG during the last 300 steps, achieving an optimal balance among generation speed, quality, diversity, and keyframe error.
See Algorithm~\ref{alg:ours} for details.

\section{Evaluation}\label{eval}
\subsection{Datasets and Metrics}
Since the SMPL representation is used in the encoding and decoding processes in NeMF,
we select the Motion-X~\cite{lin2023motionx} version of HumanML3D~\cite{Guo_2022_CVPR} in our experiments. 
We converted the data into the corresponding HumanML format when training and testing the baseline models. All motion sequences were set to a fixed length of 128 frames at 30fps.
 
The evaluation metrics used in conditional motion generation tasks~\cite{Guo_2022_CVPR} are adopted. 
\textbf{Fréchet Inception Distance (FID)} measures the difference between the distribution of generated motion data and ground-truth data by encoding motions into the latent space. 
\textbf{Diversity} is designed to measure the variability of generated motion in its distribution. 
\textbf{Keyframe error} is required to measure the average distance between generated results and ground truth keyframes in the joint-space for motion in-betweening. 
To better evaluate motion smoothness and realism, we further measure: 
\begin{itemize}
    \item \textbf{Foot-Skating Ratio}  calculates the proportion of frames where foot sliding occurs relative to the total number of generated motion frames. A frame is classified as having foot sliding when both foot joints are within 5 cm of the ground while the horizontal velocity exceeds 0.2 m/s.
    \item \textbf{Peak Jerk}~\cite{barquero2024seamless} calculates the maximum jerk value across all joints during the transition motion.
    \item \textbf{AUJ}~\cite{barquero2024seamless} is calculated by summing the L1-norm discrepancies between a method's instantaneous jerk and the dataset's average jerk value. This measure serves as an overall indicator of motion smoothness.
\end{itemize}
Although sparse keyframes provide limited semantic constraints on motion, under relatively dense keyframe settings, the semantic consistency between the generated motion and the keyframes can still serve as an indicator of sampling quality. In this case, to demonstrate our model’s strong ability to preserve motion semantic information, we report the \textbf{Matching Score} and \textbf{R-precision} for reference, even though no text conditioning is used during training or testing.

\begin{table*}[!htb]
\caption{Evaluation of Motion Completion from different numbers ($K$) of random keyframes  \label{tab:random_key}}
\resizebox{\textwidth}{!}{
\renewcommand{\arraystretch}{1.3}
\begin{tabular}{lccccccccccc}
\hline
& & Matching Score$\downarrow$   & FID$\downarrow$   & \begin{tabular}[c]{@{}c@{}}R-precision\\ (Top-3)$\uparrow$\end{tabular} & Diversity$\uparrow$  & Multi-Modality$\uparrow$  & \begin{tabular}[c]{@{}c@{}}Keyframe err$\downarrow$\end{tabular} & \begin{tabular}[c]{@{}c@{}}Foot skating \\ ratio$\downarrow$\end{tabular} & Peak Jerk$\rightarrow$ &AUJ$\downarrow$\\ \hline

& Real & $4.578^{\pm 0.015}$ & $0.001^{\pm 0.000}$ &$0.742^{\pm0.005}$ &$10.544^{\pm 0.063}$ & -  & - & $0.074^{\pm 0.001}$ & $0.033^{\pm 0.000}$ & $0.179^{\pm 0.000}$ \\ \hline

\parbox[t]{1mm}{\multirow{2}{*}{\rotatebox[origin=c]{90} {R.K=1}}}

&CondMDI & $\mathbf{5.837^{\pm 0.039}}$ &$\underline{2.071^{\pm 0.101}}$ &$\mathbf{0.630^{\pm 0.011}}$          &$\underline{9.166^{\pm 0.093}}$ &$\underline{1.914^{\pm 0.078}}$&$\underline{0.386^{\pm 0.021}}$ &$\mathbf{0.049^{\pm 0.001}}$ &$\underline{0.058^{\pm 0.001}}$ &$\underline{3.153^{\pm 0.079}}$ \\

&Ours & $\underline{6.133^{\pm 0.029}}$ &$\mathbf{1.065^{\pm 0.120}}$ &$\underline{0.536^{\pm 0.008}}$ & $\mathbf{9.666^{\pm 0.244}}$ &$\mathbf{3.355^{\pm 0.212}}$ &$\mathbf{0.083^{\pm 0.002}}$ &$\underline{0.066^{\pm 0.002}}$ &$\mathbf{0.035^{\pm 0.002}}$ &$\mathbf{0.841^{\pm 0.010}}$ \\ \hline

\parbox[t]{1mm}{\multirow{4}{*}{\rotatebox[origin=c]{90}{Rand. K=5}}} 
&SLERP &$5.617^{\pm 0.030}$ &$3.590^{\pm 0.160}$ &$0.625^{\pm 0.008}$ &$8.790^{\pm 0.225}$ &- &- &$0.145^{\pm 0.004}$ &$\underline{0.036^{\pm 0.009}}$ &$
3.410^{\pm 0.015}$\\

& NeMF &$5.286^{\pm 0.028}$ & $\underline{1.559^{\pm 0.007}}$ & $0.626^{\pm 0.009}$& $\underline{9.305^{\pm 0.272}}$ &$0.456^{\pm 0.129}$ &$\mathbf{0.023^{\pm 0.005}}$ &$\mathbf{0.057^{\pm 0.002}}$ &$0.021^{\pm 0.003}$ &$\underline{2.683^{\pm 0.074}}$\\

& CondMDI &$\underline{4.708^{\pm 0.047}}$ &$1.836^{\pm 0.252}$ &$\mathbf{0.712^{\pm 0.006}}$ &$9.250^{\pm 0.147}$ &$\underline{1.185^{\pm 0.159}}$ &$0.187^{\pm 0.006}$ &$\underline{0.060^{\pm 0.002}}$ &$0.057^{\pm 0.002}$ &$2.899 ^{\pm 0.035}$\\

& Ours &$\mathbf{4.399^{\pm 0.008}}$ &$\mathbf{0.484^{\pm 0.031}}$ &$\underline{0.676^{\pm 0.005}}$ &$\mathbf{10.051^{\pm 0.188}}$ & $\mathbf{1.284^{\pm 0.171}}$ &$\underline{0.111^{\pm 0.001}}$ &$0.064^{\pm0.001}$ &$\mathbf{0.035^{\pm 0.003}}$ &$\mathbf{0.902^{\pm 0.019}}$\\ \hline

\parbox[t]{1mm}{\multirow{4}{*}{\rotatebox[origin=c]{90}{Rand. K=20}}} 
&SLERP &$\underline{4.585^{\pm 0.014}}$ &$0.493^{\pm 0.009}$ &$0.710^{\pm 0.004}$ &$9.235^{\pm 0.204}$ &- &- &$0.134^{\pm 0.001}$ &$0.052^{\pm 0.009}$ &$1.810 ^{\pm 0.088}$\\

&NeMF  &$\mathbf{4.572^{\pm 0.023}}$ &$\mathbf{0.282^{\pm 0.012}}$ &$\underline{0.726^{\pm 0.004}}$ &$\underline{9.864^{\pm 0.206}}$ &$0.326^{\pm 0.069}$ &$\mathbf{0.033^{\pm 0.001}}$ &$\mathbf{0.065^{\pm0.001}}$ &$\underline{0.024^{\pm 0.002}}$ &$\underline{1.754^{\pm 0.049}}$\\

&CondMDI &$4.721^{\pm 0.065}$ &$1.259^{\pm 0.151}$ &$0.723^{\pm 0.016}$ &$9.398^{\pm0.171}$ &$\underline{0.368^{\pm0.051}}$ &$\underline{0.101^{\pm 0.001}}$ &$0.069^{\pm 0.001}$ &$0.057^{\pm 0.003}$ &$2.956^{\pm 0.065}$\\

&Ours &$4.759^{\pm0.055}$ &$\underline{0.348^{\pm 0.017}}$ &$\mathbf{{0.727^{\pm0.007}}}$ &$\mathbf{10.162^{\pm 0.220}}$ &$\mathbf{0.411^{\pm 0.062}}$ &$0.119^{\pm 0.001}$ &$\underline{0.067^{\pm 0.001}}$ &$\mathbf{0.031^{\pm 0.005}}$ &$\mathbf{0.682^{\pm 0.009}}$\\ \hline
\end{tabular}}
\end{table*}

\subsection{Motion Completion From Random Keyframes} \label{sec:ran_key_exp}
\begin{figure*}[!htb]
  \centering
  \includegraphics[width=1.0\textwidth]{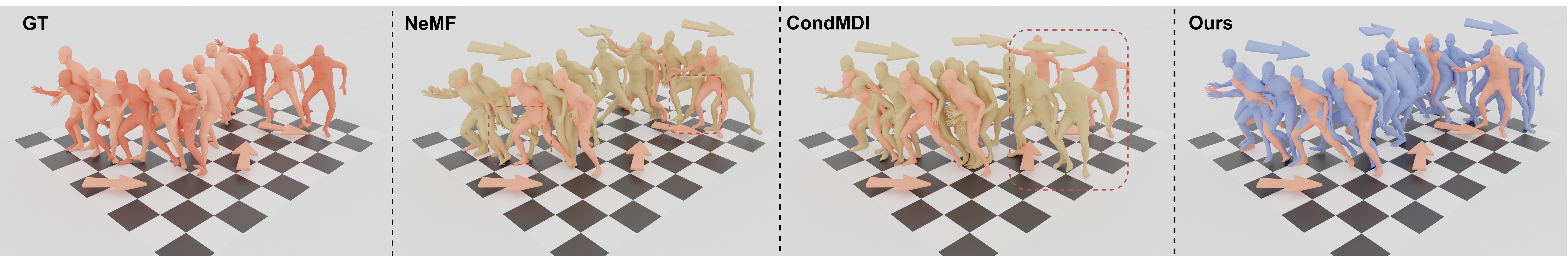}
  \caption{Qualitative comparison of state-of-the-art methods with Random
  K=5. The keyframe indices are fixed as 6, 24, 36, 90 and 110. \textbf{NeMF}: leg movements exhibit highly unnatural stepping and swinging when encountering a sharp change in direction. 
  \textbf{CondMDI}:
  gradually deviates from the direction of the keyframes and behaves with noticeable jittering in the latter half of the motion.
  }
  \label{fig:random_k=5}
  \centering
\end{figure*}

We evaluate the performance of our method and baselines on the task of sparse keyframe in-betweening. The selected baselines are all publicly released models that support random keyframe in-betweening without requiring textual conditioning: the VAE-based NeMF~\cite{NeMF} and diffusion model CondMDI~\cite{Cohan:SIGGRAPH2024:CondMDI}. Following the setup of CondMDI, we opted for the more difficult random keyframe in-betweening task instead of the conventional fixed-transition keyframe. 
We set up three scenarios: 1 keyframe (K=1), 5 keyframes (K=5, see Fig.~\ref{fig:random_k=5}), and 20 keyframes (K=20). All the keyframes are placed randomly. The test set was evaluated ten times, and each test case was tested ten times.

The results in Table~\ref{tab:random_key} indicate
our method consistently outperformed the state-of-the-art diffusion-based motion in-betweening model CondMDI on FID, Multi-Modality and Diversity with different numbers of random keyframes. This highlights that our method can generate more diverse in-between motions while preserving the quality. With a smaller number of keyframes given (K=1 and K=5), our method achieved a lower keyframe error than CondMDI.

\subsection{Motion In-betweening with Start/End Keyframes}
Generating long-term in-between motion is a challenging task.
Issues such as loss of motion details, discontinuities, and even the failure to reach target keyframe positions are highly likely to occur.
For this task, the keyframes are located at the beginning and end of the motion with K specifying the number of consecutive frame(s) to be evenly distributed to the two ends. In addition to CondMDI and NeMF, MMM~\cite{MMM} is included in the comparison since it has the capability of predicting the in-betweening frames (as tokens) without text and achieved SOTA performance on a wide range of motion generation tasks. We also introduce the two-stage transformer (TST)~\cite{Qin:SA2022} as a non-generative baseline. We used K=2 and K=8 in the experiments as K=2 is essentially sparse keyframe interpolation while K=8 is chosen since MMM tokenized every 4 frames as the minimum motion representation. The other settings and metrics of this task are the same as in Sec.~\ref{sec:ran_key_exp}.

\begin{table*}[]
 \caption{Motion In-betweening with different numbers $K$ of consecutive keyframes evenly distributed to the start and end.\label{tab:content_key}}
\resizebox{\textwidth}{!}{
\renewcommand{\arraystretch}{1.4}
\begin{tabular}{llccccccccccc}
\hline
\multicolumn{1}{c}{} &  & Matching Score$\downarrow$  & FID$\downarrow$  & \begin{tabular}[c]{@{}c@{}}R-precision\\ (Top-3)$\uparrow$\end{tabular} &Diversity$\uparrow$ &Multi-Modality$\uparrow$ & \begin{tabular}[c]{@{}c@{}}Keyframe err$\downarrow$\end{tabular} & \begin{tabular}[c]{@{}c@{}}Foot skating \\ ratio$\downarrow$\end{tabular} & Peak Jerk$\rightarrow$ &AUJ$\downarrow$\\ \hline
&Real &$4.578^{\pm 0.015}$ &$0.001^{\pm 0.000}$ &$0.742^{\pm0.005}$ &$10.544^{\pm 0.063}$ &- &- &$0.074^{\pm 0.001}$ &$0.033^{\pm 0.000}$ &$0.179^{\pm 0.000}$ \\ \hline
\parbox[t]{1mm}{\multirow{6}{*}{\rotatebox[origin=c]{90}{Start/End K=2}}} 
&SLERP &$6.559^{\pm 0.019}$ &$8.626^{\pm0.000}$ &$0.468^{\pm0.011}$ &$7.745^{\pm0.171}$ &- &- &$\underline{0.066^{\pm 0.000}}$ &$0.002^{\pm 0.000}$ &$4.129^{\pm 0.000}$ \\
&TST &$6.757^{\pm0.028}$ & $9.367^{\pm 0.042}$ &$0.532^{\pm 0.019}$ &$6.899^{\pm0.056}$ & - &$0.091^{\pm0.004}$ &$0.096^{\pm0.002}$ &$0.308^{\pm0.009}$ &$4.668^{\pm0.093}$\\ 
&NeMF &$6.204^{\pm 0.034}$ &$\underline{6.343^{\pm0.051}}$ &$0.469^{\pm 0.013}$ &$8.104^{\pm0.104}$ &$0.535^{\pm0.056}$ &$\mathbf{0.022^{\pm0.000}}$ &$\mathbf{0.046^{\pm0.002}}$ &$\underline{0.009^{\pm 0.003}}$ &$\underline{3.408^{\pm 0.049}}$ \\
&CondMDI &$\underline{5.710^{\pm 0.049}}$ &$8.074^{\pm 0.336}$ &$\underline{0.606^{\pm 0.018}}$ &$\underline{9.119^{\pm 0.143}}$ &$\underline{1.914^{\pm 0.078}}$ &$0.461^{\pm 0.005}$ &$0.084^{\pm 0.001}$ &$0.145^{\pm 0.003}$ &$4.064^{\pm 0.102}$ \\
&MMM &- &- &- &- &- &- &- &- &-\\
&Ours &$\mathbf{5.485^{\pm 0.049}}$ &$\mathbf{0.524^{\pm 0.047}}$ &$\mathbf{0.616^{\pm 0.007}}$ &$\mathbf{9.746^{\pm 0.136}}$ &$\mathbf{2.252^{\pm 0.264}}$ &$\underline{0.081^{\pm 0.001}}$ &$0.073^{\pm 0.001}$ &$\mathbf{0.032^{\pm 0.004}}$ &$\mathbf{0.790^{\pm 0.010}}$ \\ \hline
\parbox[t]{1mm}{\multirow{6}{*}{\rotatebox[origin=c]{90}{Start/End K=8}}} 
&SLERP &$6.418^{\pm 0.021}$ &$7.422^{\pm 0.0000}$ &$0.516^{\pm 0.011}$ &$7.925^{\pm0.142}$ &- &- &$0.066^{\pm 0.0007}$ &$0.0574^{\pm 0.000}$  &$4.031^{\pm 0.049}$ \\
&TST &$6.956^{\pm0.017}$ & $9.243^{\pm 0.043}$ &$0.542^{\pm 0.011}$ &$6.260^{\pm0.088}$ &-  &$\underline{0.068^{\pm0.004}}$ &$0.087^{\pm0.003}$      &$0.404^{\pm 0.013}$ &$4.931^{\pm 0.059}$ \\
&NeMF &$6.251^{\pm0.017}$ & $5.409^{\pm 0.043}$ &$0.516^{\pm 0.011}$ &$8.354^{\pm0.155}$ & $0.586^{\pm0.148}$ &$\mathbf{0.023^{\pm0.001}}$ &$\mathbf{0.046^{\pm0.001}}$                &$\underline{0.021^{\pm 0.002}}$ &$\underline{2.683^{\pm 0.062}}$ \\
&CondMDI  &$\mathbf{4.744^{\pm0.037}}$ &$8.675^{\pm 0.246}$ &$\underline{0.612^{\pm 0.007}}$ &$7.727^{\pm 0.083}$ &$1.581^{\pm 0.041}$ &$0.332^{\pm0.030}$ & $0.073^{\pm 0.001}$ &$0.130^{\pm 0.003}$  &$3.361^{\pm 0.070}$ \\
&MMM &$8.007^{\pm 0.017}$ &$\underline{3.573^{\pm 0.080}}$ &$0.347^{\pm 0.014}$ &$\mathbf{10.829^{\pm 0.063}}$ &$\mathbf{2.571^{\pm0.348}}$ &$0.630^{\pm 0.003}$ &$\underline{0.057^{\pm0.000}}$ &$0.482^{\pm 0.072}$  &$3.916^{\pm 0.149}$ \\
&Ours  & $\underline{5.548^{\pm0.037}}$ & $\mathbf{0.656^{\pm 0.039}}$ & $\mathbf{0.620^{\pm0.010}}$ & $\underline{9.411^{\pm0.159}}$ & $\underline{1.627^{\pm0.348}}$ &$0.093^{\pm0.001}$ &$0.069^{\pm0.001}$ &$\mathbf{0.033^{\pm 0.004}}$ &$\mathbf{1.088^{\pm 0.010}}$  \\ \hline
\end{tabular}}
\end{table*}

\begin{figure*}[htb]
  \centering
  \includegraphics[width=1.0\textwidth]{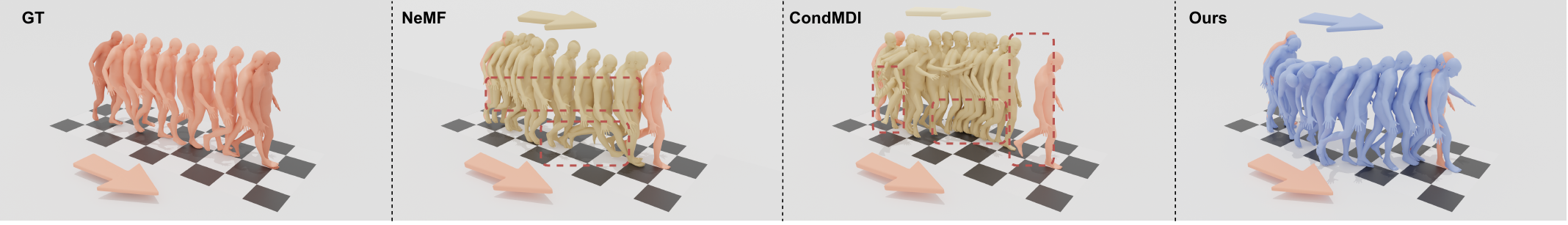}
  \caption{Qualitative comparison of state-of-the-art methods with Start/End K=2. \textbf{NeMF}: The movement of the upper body loses most of the detail, and the arms remain relatively still throughout. The feet still exhibit unnaturally large strides. \textbf{CondMDI}: It has high keyframe error and exhibits long-distance displacement at the end of motion.} 
  \label{fig:content_k=1}
  \centering
\end{figure*}

From Table~\ref{tab:content_key}, our method demonstrated superiority on most of the quality and motion diversity-related metrics while achieving a comparable performance on keyframe error and foot skating ratio with NeMF when minimum keyframe information is given (i.e K=2). In particular, our method achieved an extremely low FID when compared with other methods on different numbers of content frames. An example is illustrated in Fig.~\ref{fig:content_k=1}. Our method achieved comparable keyframe error with NeMF, which is much smaller than other generative models MMM and CondMDI. Our result also shows a smoother walking motion with more uniform stride length over NeMF and CondMDI, which was also indicated in the JERK metrics quantitatively. Readers are referred to the video demo for more results. 

\begin{figure*}[!htb]
  \centering
  \includegraphics[width=1.0\textwidth]{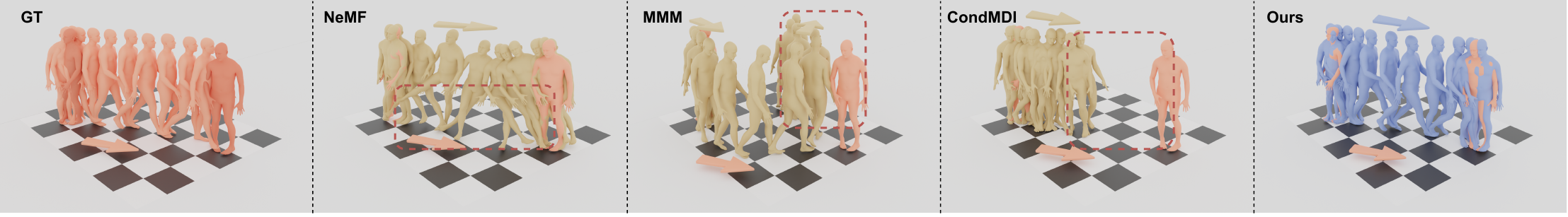}
  \caption{Qualitative comparison of state-of-the-art methods with Start/End K=8. \textbf{NeMF}: completely distorted, with severe body swaying and translation. \textbf{MMM}: global location information is lost and walked in the opposite direction.  \textbf{CondMDI}: similar to the results when K=2; failed to reach the target position.}
  \label{fig:content_k=4}
  \centering
\end{figure*}

With K=8 (Fig.~\ref{fig:content_k=4}), MMM performs well on diversity and CondMDI achieves the best results on semantic meaning. Nevertheless, our method demonstrated a more balanced performance with the best performance on FID and the second-best performance on 4 other metrics. For MMM, the character failed to reach the end keyframe because only root velocity is encoded in MMM, and it failed to recover the global root trajectory due to error accumulation. TST lags behind generative models in terms of generation quality and is unable to generate diverse results.

\section{Ablation Study and Hyper-parameter Experiment}
\label{sec:abla}
\subsection{Using INR Over Other Representations}
\label{sec:abla_inr}

To evaluate the effectiveness of INR, we replaced the INR VAE of our model with the transformer VAE from MLD~\cite{mld} and the 1D-CNN VQVAE from MMM~\cite{MMM}.
Compared to the other representations, using INR resulted in better motion quality across different settings, as presented in Table~\ref{tab:abla_inr}. 

\begin{table}[]
\caption{Ablation study of motion representation\label{tab:abla_inr}}

\resizebox{\linewidth}{!}{
\renewcommand{\arraystretch}{1.2}
\begin{tabular}{clcccc}
\hline
\multicolumn{1}{l}{} &  & FID$\downarrow$ & \begin{tabular}[c]{@{}c@{}} R\_precision.\\(Top-3)$\uparrow$ \end{tabular}& Div.$\uparrow$ & Keyframe Err.$\downarrow$ \\ \hline
\multirow{6}{*}{Rand K=5} & MLD & $1.122$ & $0.619$& $9.153$ & $\mathbf{0.311}$ \\
& MMM & $4.769$ & $0.562$& $8.047$ & $0.451$ \\
& Ours & $\mathbf{0.892}$& $\mathbf{0.638}$ & $\mathbf{9.525}$ & $0.337$ \\ \cline{2-6} 
& MLD w.IMG & $1.232$ & $0.613$ & $9.079$ & $0.229$ \\
& MMM w.IMG & $0.801$ & $0.628$ & $9.499$ & $0.176$ \\
& Ours w.IMG &$\mathbf{0.484}$&$\mathbf{0.676}$ &$\mathbf{10.051}$  &$\mathbf{0.111}$  \\ \hline
\multirow{6}{*}{S./E. K=8} & MLD & $2.209$ & $0.537$ & $8.643$ & $\mathbf{0.234}$ \\
& MMM & $6.953$ & $0.413$ & $7.153$ & $0.518$ \\
& Ours & $\mathbf{1.491}$ & $\mathbf{0.572}$ & $\mathbf{9.113}$ & $0.257$ \\ \cline{2-6} 
& MLD w.IMG & $2.274$ & $0.534$ & $8.743$ & $0.203$ \\
& MMM w.IMG & $3.253$ & $0.472$ & $8.683$ & $0.214$ \\
& Ours w.IMG & $\mathbf{0.656}$ & $\mathbf{0.602}$ & $\mathbf{9.411}$ & $\mathbf{0.092}$ \\ \hline
\end{tabular}}
\label{tab:abla_inr}
\end{table}

\subsection{Evaluation of IMG}

\begin{figure*}[htb]
  \centering
  \includegraphics[width=1\textwidth]{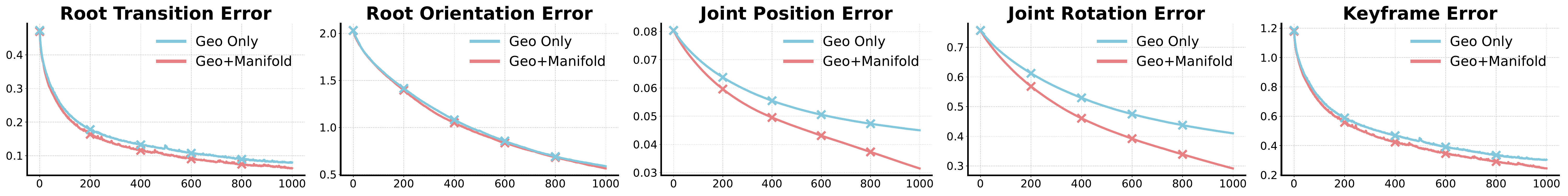}
  \caption{Various pose-level errors - Ablation study without diffusion.} 
  \label{fig:abl_curves}
  \centering
\end{figure*}

To further verify the effectiveness of our IMG optimization, we conducted a two-stage ablation study using the test-set from HumanML3D. In the first phase, we randomly generated a set of Gaussian noise and directly applied IMG with two different options to perform gradient descent on the manifold. The first option uses only geometric loss, while the second one includes both geometric and manifold losses. This ensures the results are not affected by the denoising process in the diffusion model. The keyframe number is set to 5 and they are placed evenly across the motion. 
Since the step of predicting $\hat{z}_{0}$ from $\hat{z}_{t}$ transition is removed, both options are directly optimizing the current $\hat{z}_{t}$. Fig.~\ref{fig:abl_curves} illustrates the variation in different errors and highlights the manifold constraint significantly accelerated convergence, especially in joint rotational errors.

In the second stage, we reintroduce the latent diffusion model and conduct quantitative evaluations on our trained model using various diffusion editing methods, including DNO and DPS. For DNO, the subsequent optimization runs for 300 steps through direct 10-step DDIM sampling without any inversion.
We also evaluate the effectiveness of IMG under different constraint formulations. IMG(Only Geo) constrains the sampling process solely based on keyframe reconstruction error. IMG (Slerp), on the other hand, performs Slerp between the generated motion at each denoising step and the keyframes to obtain a reference motion, and then uses the error between the reference motion and the generated motion as the constraint for guidance. IMG and IMG (300) follow the standard guidance procedure, with the latter applying optimization only during the final 300 steps.
Two typical scenarios are chosen - Random 5 Keyframes and 2 Start/End Keyframes in-betweening, and the experimental setup is identical to the one in Sec.~\ref{eval}.
The results are presented in 
Table~\ref{tab:eval_img}. Compared to other baselines, our optimization algorithm achieves the best performance across all metrics except for foot sliding. Notably, the results with the added manifold correction term significantly reduce keyframe error and the FID of the generated motions. Even when applied only in the last 300 steps, our full optimization algorithm maintains superior performance while preserving the original diversity of the generated motion.

\subsection{Hyper-parameter Experiment}
We furthermore validated the sampling results of IMG with different numbers of optimization steps and different Gaussian kernel sizes. These experiments show that IMG achieves low keyframe error with fewer sampling steps while still maintaining high diversity, demonstrating excellent reliability and convenience. We also tested the trade-off between keyframe constraints and sampling quality and diversity. As the constraint scale increases, the quality and diversity of generated actions decrease significantly. The influence of CFG scale is also evaluated on sampling performance. The results show that when the CFG scale is within a small range, the motion quality remains largely unchanged, while the keyframe error is slightly reduced. However, as the CFG scale increases significantly, the sampling results exhibit noticeable degradation. All detailed results are shown in Table~\ref{tab:ab}.

\begin{table*}[htbp]
\caption{Quantitative comparison of our IMG with other diffusion sampling guidance methods.}
\label{tab:eval_img}
\resizebox{\textwidth}{!}{
\renewcommand{\arraystretch}{1.3}
\begin{tabular}{lcccccccc}
\hline 
\multicolumn{1}{c}{} &  & Matching Score$\downarrow$ & \begin{tabular}[c]{@{}c@{}} R\_precision\\(Top-3)$\uparrow$ \end{tabular} & FID$\downarrow$ & Diversity$\uparrow$ & Multi\_Modal$\uparrow$ & KeyFrame\_Error$\downarrow$ &\begin{tabular}[c]{@{}c@{}} Foot\_Skating\\Ratio$\downarrow$\end{tabular} \\  \hline 
Real & \multicolumn{1}{l}{}  & $4.578 ^{\pm 0.015}$ & $0.742^{\pm 0.005}$ & $0.001^{\pm 0.000}$ &$ 10.544^{\pm 0.063}$ &-  & - & $0.074^{\pm 0.001}$ \\ \hline 
Origin (CFG1.2) & \multicolumn{1}{l}{}  & $5.191^{\pm 0.026}$ & $0.624^{\pm 0.011}$ & $0.888^{\pm 0.056}$ & $9.521^{\pm 0.178}$ & $\mathbf{1.400^{\pm 0.070}}$ & $0.327^{\pm 0.009}$ & $0.071^{\pm 0.001}$ \\ 
DNO & \multicolumn{1}{l}{}  & $5.117^{\pm 0.042}$ & $0.654^{\pm 0.011}$ & $0.969^{\pm 0.031}$ & $\underline{9.828^{\pm 0.165}}$ & $0.219^{\pm 0.003}$ & $0.225^{\pm 0.005}$ & $\underline{0.065^{\pm 0.002}}$ \\
DPS & \multicolumn{1}{l}{}  & $5.102^{\pm 0.041}$ & $0.635^{\pm 0.016}$ & $0.781^{\pm 0.055}$ & $9.603^{\pm 0.192}$ & $1.270^{\pm 0.078}$ & $0.185^{\pm 0.003}$ & $0.071^{\pm 0.001}$ \\
Ours(Only Geo) &  &$5.098^{\pm 0.033}$ & $0.655^{\pm 0.006}$ & $0.596^{\pm 0.028}$ & $9.454^{\pm 0.201}$ & $1.103^{\pm 0.147}$ & $0.129^{\pm 0.007}$ & $0.070^{\pm 0.001}$ \\ 
Ours(Slerp) & \multicolumn{1}{l}{}  & $5.097^{\pm0.038}$ & $0.652^{\pm0.014}$ & $0.766^{\pm0.051}$ & $9.540^{\pm0.189}$ & $1.236^{\pm0.101}$ & $0.150^{\pm0.002}$ & $0.076^{\pm0.002}$ \\ 
Ours & \multicolumn{1}{l}{} & $\underline{4.872^{\pm 0.046}}$ & $\underline{0.657^{\pm 0.013}}$ & $\mathbf{0.450^{\pm 0.042}}$ & $9.519^{\pm 0.138}$ & $1.090^{\pm 0.098}$ &  $\mathbf{0.098^{\pm 0.003}}$ & $0.068^{\pm 0.001}$ \\ 
Ours(T=300) & \multirow{-7}{*}{Random K=5} & $\mathbf{4.399^{\pm 0.008}}$ &$ \mathbf{0.676^{\pm 0.005}}$ & $\underline{0.484^{\pm 0.031}}$ & $\mathbf{10.050^{\pm 0.189}}$ & $\underline{1.284^{\pm 0.171}}$ & $\underline{0.111^{\pm 0.001}}$ & $\mathbf{0.064^{\pm 0.001}}$ \\ \hline 
Origin (CFG1.2) & \multicolumn{1}{l}{} & $5.796^{\pm 0.032}$ & $0.573^{\pm 0.026}$ & $1.465^{\pm 0.129}$ & $8.894^{\pm 0.201}$ & $\mathbf{1.844^{\pm 0.087}}$ & $0.347^{\pm 0.006}$ & $\underline{0.068^{\pm 0.002}}$ \\ 
DNO & \multicolumn{1}{l}{}  & $5.788^{\pm 0.031}$ & $0.572^{\pm 0.011}$ & $2.077^{\pm 0.096}$ & $9.309^{\pm 0.164}$ & $0.341^{\pm 0.045}$ & $0.229^{\pm 0.003}$ & $\mathbf{0.065^{\pm 0.006}}$ \\
DPS & \multicolumn{1}{l}{} & $5.737^{\pm 0.071}$ & $0.569^{\pm 0.007}$ & $1.495^{\pm 0.031}$ & $8.942^{\pm 0.177}$ & $1.379^{\pm 0.124}$ & $0.156^{\pm 0.001}$ & $0.070^{\pm 0.002}$ \\ 
Ours(Only Geo) & \multicolumn{1}{l}{} & $5.649^{\pm 0.026}$ & $\underline{0.593^{\pm0.010}}$ & $1.305^{\pm 0.066}$ & $9.306^{\pm 0.195}$ & $1.638^{\pm 0.163}$ & $\underline{0.098^{\pm 0.007}}$ & $0.071^{\pm 0.001}$ \\
Ours(Slerp) & \multicolumn{1}{l}{} & $5.746^{\pm0.018}$ & $0.573^{\pm0.007}$ & $1.787^{\pm0.168}$ & $9.020^{\pm0.162}$ & $\underline{1.718^{\pm0.162}}$ & $0.168^{\pm0.002}$ & $0.078^{\pm0.001}$ \\ 
Ours & \multicolumn{1}{l}{}  & $\underline{5.582^{\pm 0.070}}$ & $0.592^{\pm 0.020}$ & $\underline{0.669^{\pm 0.029}}$ & $\mathbf{9.683^{\pm 0.066}}$ & $1.145^{\pm 0.172}$ & $\mathbf{0.091^{\pm 0.019}}$ & $0.072^{\pm 0.003}$  \\ 
Ours(T=300) & \multirow{-7}{*}{Start/End K=8} & $\mathbf{5.548^{\pm 0.037}}$ & $\mathbf{0.602^{\pm 0.010}}$ & $\mathbf{0.656^{\pm 0.039}}$ &$\underline{9.411^{\pm 0.159}}$ & $1.627^{\pm 0.348}$ & $0.109^{\pm 0.017}$ &  $0.069^{\pm  0.001}$  \\ \hline 
\end{tabular}}
\begin{tablenotes}
    \item[1]By default, DPS optimizes at each step.
\end{tablenotes}
\end{table*}

\begin{table}[]
\caption{Ablation study on Optimization Steps, Gaussian filter size and Geometric Constraint Scale}
\renewcommand{\arraystretch}{1.2}
\resizebox{\linewidth}{!}{
\begin{tabular}{lcccc}
\hline
& FID$\downarrow$ & Div. $\uparrow$& MM$\uparrow$ & Keyframe Err.$\downarrow$  \\ \hline
Optim Steps &  &  &  &  \\ \hline
$T=1000$ & $\mathbf{0.450}$ & $\mathbf{9.419}$ & $0.890$ & $\mathbf{0.101}$ \\
$T=500$  & $0.452$ & 9.411 & 1.000 & 0.103 \\
$T=300$  & 0.457 & 9.412 & 1.095 & 0.107  \\
$T=100$ & 0.492 & 9.397 & $\mathbf{1.112}$ & 0.131 \\ \hline
Kernal Size &  &  &  &   \\ \hline
$w=0$ (Geo Only) & 0.569 & 9.329 & $\mathbf{1.131}$ & 0.116\\
$w=1$ & 0.479 & 9.390 & 1.059 & 0.109 \\
$w=5$ & 0.475 & $9.404$& 1.051 & $0.107$ \\
$w=9$ & $\mathbf{0.457}$ & $\mathbf{9.412}$ & $1.095$ & $0.107$ \\
$w=13$ & 0.458 & $9.404$ & 1.043 & $\mathbf{0.106}$ \\ \hline
Geometric Constraint Scale &  &  &  &   \\ \hline
Origin & 0.888 & $\mathbf{9.521}$ & $\mathbf{1.400}$ & 0.327\\
$s_{geo}=1$ & $\mathbf{0.596}$ & 9.453 & 1.103 & 0.129 \\
$s_{geo}=5$ & 0.603 & $9.340$& 1.093 & $0.088$ \\
$s_{geo}=10$ & $0.617$ & $9.262$ & $0.991$ & $\mathbf{0.086}$ \\\hline
Class Free Guidance &  &  &  &   \\ \hline
CFG $=1.0$ &  $\mathbf{0.448}$ & $\mathbf{9.564}$ & 1.012 & 0.108\\
CFG $=1.2$  & 0.450 & 9.519 & 1.090 & $\mathbf{0.096}$ \\
CFG $=1.5$  & 0.578 & 9.511 & 1.040 & 0.098 \\
CFG $=2.0$  & 1.173 & 9.298 & $\mathbf{1.179}$ & 0.119 \\\hline
\end{tabular}}
\label{tab:ab}
\end{table}

\section{Unconditional Path Following}

To show the versatility of our model and sampling optimization method, we also examined the model's performance on root-path following. Here, no additional conditionings, including keyframe, skeleton joint, and textual semantics, are provided throughout the sampling process. We compare our model with CondMDI which shares the same training setup and metrics (except replacing keyframe error with trajectory error) with the previous experiment. 
Our model achieved better motion quality without any additional training, see Table~\ref{tab:root}.

\begin{table}[htbp]
\caption{Results on Unconditional trajectory following.}\label{tab:root}
\renewcommand{\arraystretch}{1.3}
\resizebox{0.48\textwidth}{!}{
\begin{tabular}{lcccc}
\hline
& FID$\downarrow$ & Div.$\uparrow$ & MM$\uparrow$ & Traj\_Err.$\downarrow$ \\ \hline
NeMF  & $23.297^{\pm 0.410}$ &$6.011^{\pm 0.087}$ & $\mathbf{4.590^{\pm 0.172}}$ & $0.139^{\pm 0.004}$ \\
CondMDI  & $1.638^{\pm0.151}$ & $9.197^{\pm0.080}$ & $1.314^{\pm0.104}$& $0.054^{\pm0.001}$ \\
Ours  & $\mathbf{1.008^{\pm0.048}}$ & $\mathbf{9.281^{\pm0.163}}$ & $1.439 ^{\pm0.096}$ & $\mathbf{0.052^{\pm0.001}}$  \\ \hline

\end{tabular}
}
\end{table}

\section{Cross-Dataset Performance}
To evaluate the generalization ability, 
we tested our model on the AIST++ \cite{li2021ai} while trained only on the HumanML3D. Table \ref{tab:cross_dataset} indicates that our model maintains strong performance even on unseen data distributions. This highlights the robust generalization ability of our model.

\begin{table*}[htbp!]
\caption{Quantitative comparison of our pipeline with another baseline on the cross-dataset.}
\label{tab:cross_dataset}
\resizebox{\textwidth}{!}{
\renewcommand{\arraystretch}{1.3}

\begin{tabular}{clccccccc}
\hline
\multicolumn{1}{l}{} &  & FID$\downarrow$ & Div.$\uparrow$ & MM$\uparrow$  & Key. Err.$\downarrow$& Foot\_Skating$\downarrow$ & PeakJerk$\rightarrow$ & AUJ$\downarrow$ \\ \hline
\multirow{4}{*}{S./E. K=8} & NeMF & $4.204^{\pm0.084}$ & $5.733^{\pm0.259}$ & $0.799^{\pm0.022}$ & $\mathbf{0.030^{\pm0.008}}$ & $\mathbf{0.024^{\pm0.001}}$ & $0.025^{\pm0.002}$  & $2.712^{\pm0.050}$\\
 & CondMDI & $10.942^{\pm0.349}$ & $5.041^{\pm0.204}$ & $1.904^{\pm0.068}$ & $1.921^{\pm0.172}$ & $0.063^{\pm0.022}$ & $0.536^{\pm0.013}$  & $8.917^{\pm0.233}$ \\
 & MMM & $59.393^{\pm0.613}$ & $\mathbf{8.064^{\pm0.137}}$ & $\mathbf{6.386^{\pm0.692}}$ & $3.662^{\pm0.054}$ & $0.062^{\pm0.003}$ & $0.304^{\pm0.009}$ & $4.431^{\pm0.194}$ \\
 & Ours &$\mathbf{3.131^{\pm0.052}}$  & $5.631^{\pm0.091}$ & $2.493^{\pm0.037}$ & $0.125^{\pm0.012}$ & $0.059^{\pm0.018}$ & $0.048^{\pm0.002}$ & $\mathbf{0.706^{\pm0.082}}$ \\ \hline
\end{tabular}}

\end{table*}

\section{Computational Efficiency}
\begin{figure}[!htb]
  \centering
  \includegraphics[width=0.48\textwidth]{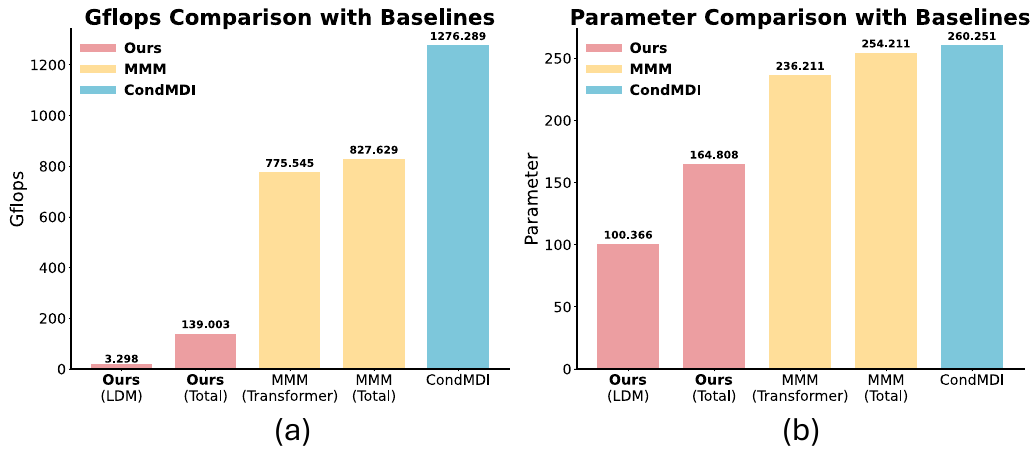}
  \caption{The comparison of model size and computational cost.}
  \label{fig:cal}
  \centering
\end{figure}

Our DiT has linear complexity when computing owing to the cross-attention structure. The query input of this cross-attention structure comes solely from the 1D INR parameter, rather than taking the entire motion sequence as in self-attention models. In contrast, the typical temporal-latent DiT's computational complexity would be quadratic in the generated sequence length. This characteristic significantly reduces the computational cost of our model.
We calculate the number of parameters and floating-point operations for our model and other baseline models. The results are shown in Fig.~\ref{fig:cal}.

We also measured the sampling speed of both the baseline and our model with a batch size of 32 and 5 keyframes. All tests were conducted in an environment with one NVIDIA RTX 4090 in CUDA 12.4. CondMDI uses the default 1000 DDPM sampling steps, while NeMF uses the default 800 optimization steps. The configuration for DNO is the same as in the ablation study, with 300 optimization steps, each consisting of 10 DDIM sampling steps. Since our sampling optimization is flexible and can be adjusted according to different requirements, we compare several settings: without IMG, 100-step IMG sampling, 300-step IMG sampling, and the full 1000-step sampling. We report the average results over ten runs, and all results are listed in Tab.~\ref{tab:sam_time}.

\begin{table*}[!htb]
\caption{Inference time comparison with other models}
\label{tab:sam_time}
\resizebox{\textwidth}{!}{
\renewcommand{\arraystretch}{1.3}
\begin{tabular}{lccccccc}
\hline
        & CondMDI    & NeMF        & DNO        & Ours (w/o IMG)      & Ours (IMG100) & Ours (IMG300) & Ours (IMG1000) \\ \hline
Time(s) &$44.72^{\pm0.86}$ & $118.97^{\pm3.59}$ & $88.66^{\pm1.93}$ &$6.96^{\pm0.22}$ & $18.63^{\pm0.87}$   &$36.73^{\pm0.33}$   & $97.61^{\pm2.46}$    \\ \hline
\end{tabular}}
\end{table*}

\begin{figure*}[h]
  \centering
  \includegraphics[width=1.0\textwidth]{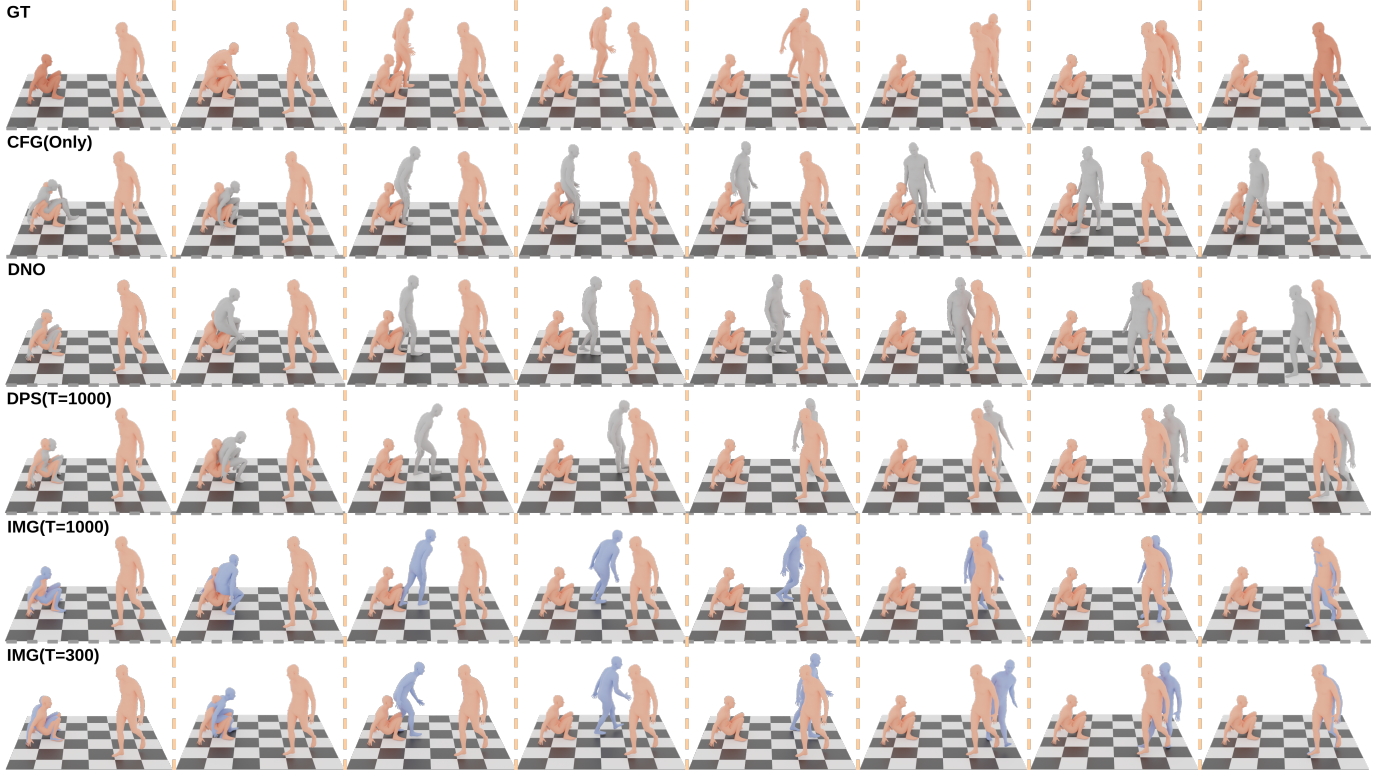}
  \caption{Ablation study of IMG with Start/End K=2. The motion index is 000066. \textbf{CFG(Only)}: The movement is generated only with class-free guidance. It captured the main features of keyframes and generated a reasonable and smooth transition, but it couldn't restore the global position perfectly.
  \textbf{DNO}: DNO presents the largest keyframe error among all optimization algorithms.
  \textbf{DPS(T=1000)}: The DPS is implemented through the whole DDIM sampling process. It improves the keyframe's global transform error. \textbf{IMG(T=1000)}: Compared to the DPS, it has better keyframe accuracy. \textbf{IMG(T=300)}: The IMG is only implemented at the last 300 steps. The result still remains excellent performances} 
  \label{fig:abla_vis}
  \centering
  \vspace{-0.5em}
\end{figure*}
\section{Conclusion and Discussion}

We have proposed a new model to address the task of motion in-betweening given very sparse keyframes. Our approach uses a diffusion model in the motion implicit neural representation space, which enables sampling continuous and smooth motion without losing fine details. We have introduced a new guidance approach that retains the strong diversity of the generative model while also minimizing keyframe error. Our model achieves state-of-the-art generation performance in the most challenging settings with very sparse keyframes. 

\paragraph{Limitations and Future Work.}
Although our pipeline achieves SOTA performance in most motion in-betweening and motion completion metrics, there is still room for improvement in keyframe error and trajectory diversity. Furthermore, while the diffusion model performs sampling in an implicit low-dimensional space, which accelerates the process, the introduction of IMG still results in a relatively lengthy sampling procedure. Moreover, we found that motion INR can serve as a strong motion prior in our experiments. We will explore motion-refinement tasks using our pipeline.

{
\bibliographystyle{IEEEtran}
\bibliography{ref}
}

\begin{IEEEbiography}[{\includegraphics[width=1in,height=1.25in,clip,keepaspectratio]{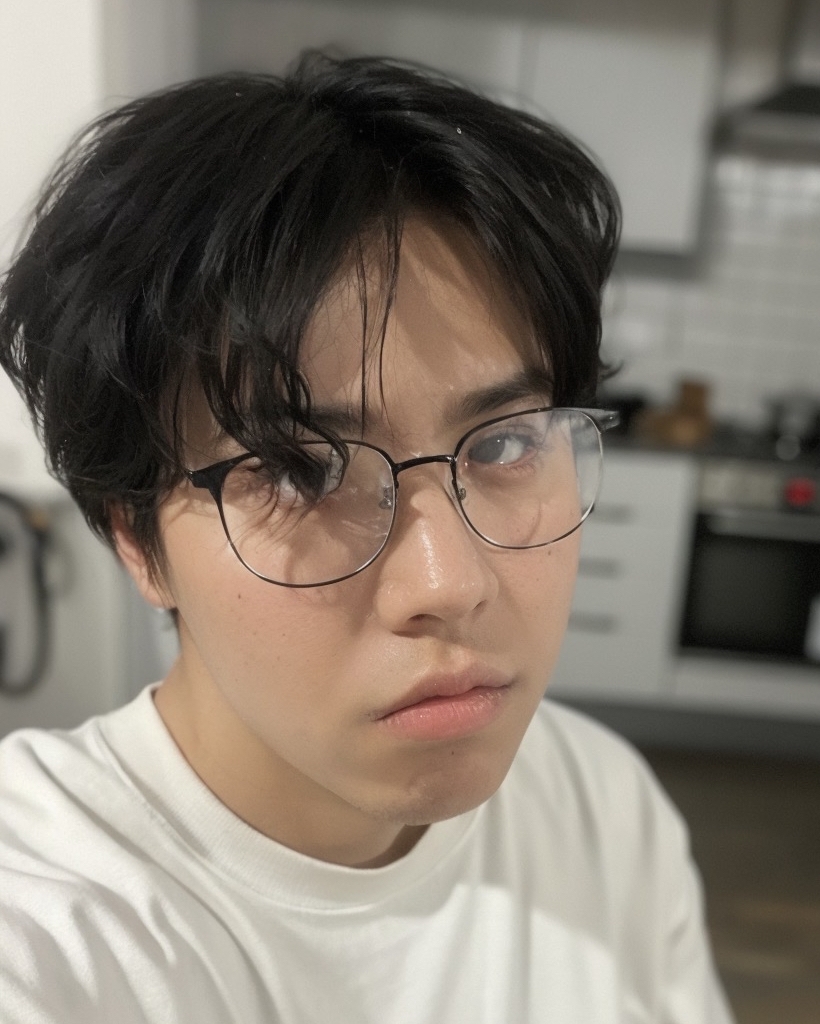}}]{Shiyu Fan} received his bachelor’s degree in Electronic and Information Engineering from Nanjing University of Science and Technology, China, and his MSc degree in Computer Science from George Washington University. 

He is currently a fourth-year Ph.D. candidate at the University of Glasgow. His primary research interests include motion control, motion generation, and embodied intelligence. He is also highly interested in generative models and other areas of computer graphics.
\end{IEEEbiography}

\begin{IEEEbiography}[{\includegraphics[width=1in,height=1.25in,clip,keepaspectratio]{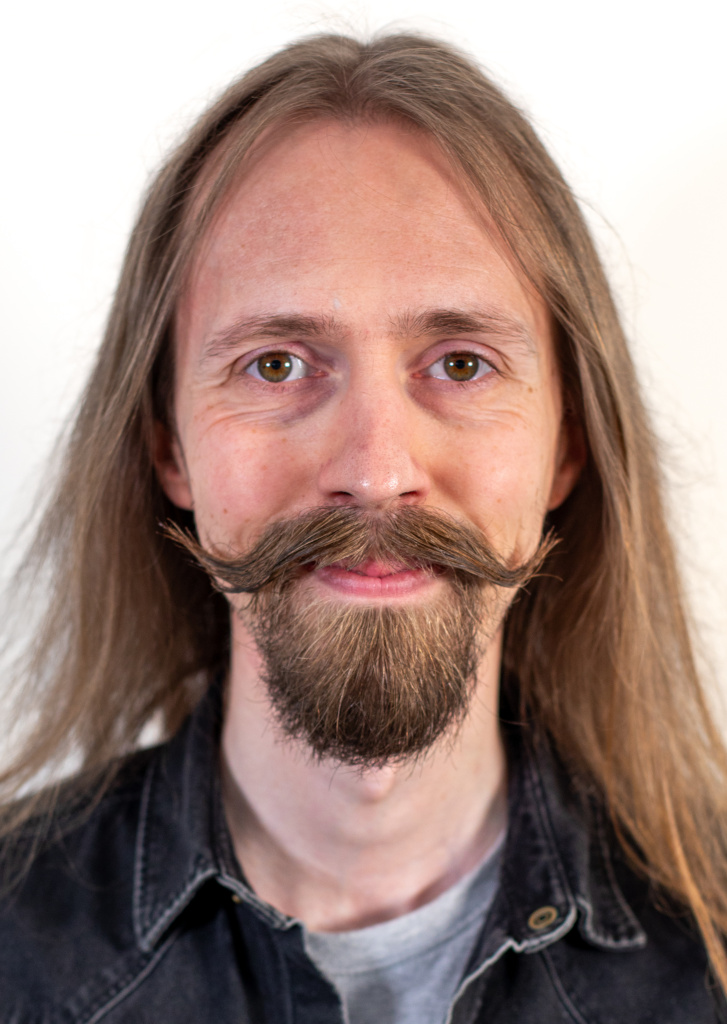}}]{Paul Henderson} received a BA in Mathematics from the University of Cambridge in 2009, MSc in Artificial Intelligence from the University of Edinburgh in 2010, and PhD in Informatics from the University of Edinburgh in 2018.
Since 2022 he is a Lecturer (Assistant Professor) in Machine Learning at the University of Glasgow, where he specializes in generative AI for visual data, probabilistic machine learning, and 3D computer vision, including applications in the physical sciences and healthcare.
\end{IEEEbiography}

\begin{IEEEbiography}[{\includegraphics[width=1in,height=1.25in,clip,keepaspectratio]{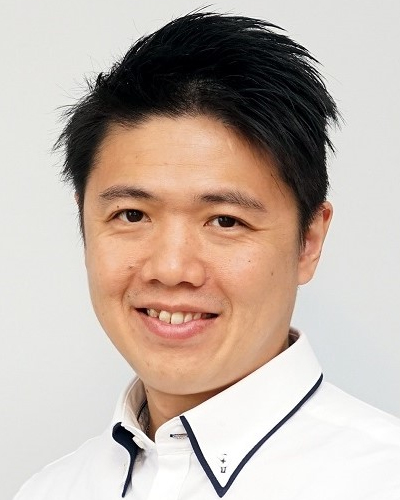}}]{Edmond S. L. Ho} received the BSc degree in computer science from the Hong Kong Baptist University in 2003, the MPhil degree in computer science from the City University of Hong Kong in 2006, and the PhD degree in informatics from the University of Edinburgh, UK in 2011. 

He is currently a Senior Lecturer (Associate Professor) in the School of Computing Science at the University of Glasgow, UK. 
He was an Associate Professor (2016-2022) in the Department of Computer and Information Sciences 
at Northumbria University, UK and a Research Assistant Professor in the Department of Computer Science at Hong Kong Baptist University (2011-2016). His research interests include Computer Graphics, Computer Vision, and Biomedical Engineering, and Machine Learning.
\end{IEEEbiography}

\vfill

\end{document}